# Probing the Locally Generated Even and Odd Order Nonlinearity in Y-Ba-Cu-O and Tl-Ba-Ca-Cu-O (2212) Microwave Resonators around $T_C$

Brooke M. Jeries, Sean R. Cratty, and S. K. Remillard, *Member, IEEE*
Physics Department, Hope College, Holland, MI 49423, USA

*Abstract*— Spatial scanning of the synchronously generated $2^{nd}$ and $3^{rd}$ order intermodulation distortion in superconducting resonators uncovers local nonlinearity hot spots, and possible time reversal symmetry breaking, using a simple probe fashioned from coaxial cable.  It is clear that even and odd order nonlinearity in these samples do not share the same physical origins, as their temperature and static magnetic field dependences are quite different.  $2^{nd}$ order intermodulation distortion (IMD) remains strong in these measurements as the temperature continues to drop below $T_C$ to 77K even though the $3^{rd}$ order peaks near $T_C$ and becomes smaller at lower temperature as predicted by the nonlinear Meissner effect. Both $YBa_2Cu_3O_7$ and $Tl_2Ba_2CaCu_2O_8$ resonators of the same structure exhibit similar temperature dependence in the $2^{nd}$ order with $2^{nd}$ order remaining high at lower temperature.  The $YBa_2Cu_3O_7$ sample has lower $3^{rd}$ order IMD with a pronounced peak at $T_C$.

## I. Introduction

Research in the nonlinearity of high temperature superconductors (HTS) using microwave signals serves at least two objectives.  First is the scientific goal of understanding superconductor electrodynamics.  For example, the anisotropic d-wave electron pairing symmetry of $YBa_2Cu_3O_7$ can be examined by $3^{rd}$ order microwave nonlinear distortion[1,2], as can that of $MgB_2$ [3,4].  Second is exploitation of HTS in technological advancements of both evolutionary and revolutionary scope.  Through highly selective low noise frequency filters, HTS thin films have the potential to threaten the technology paradigm of wireless [5,6] and space-based telecommunications[7,8], and will improve the resolution of magnetic resonance imaging[9].  Microwave resonators and filters manufactured for commercial deployment serve as the samples in our research.

The workhorse materials in these application successes have been $YBa_2Cu_3O_7$ (YBCO) and $Tl_2Ba_2CaCu_2O_8$ (TBCCO) thin films.  We have been studying the nonlinearity of patterned resonators made from these materials on $LaAlO_3$ (LAO), MgO and sapphire substrates.  This paper reports recent results using LAO substrates providing (1) comparison between the nonlinearity of YBCO and TBCCO at high temperatures, and (2) scanning of the nonlinearity across the resonator.

Zhao, *et al.*[10] found that the double thallium layer TBCCO films with weakly pinned pancake fluxons, had higher $3^{rd}$ order nonlinearity than YBCO, suggesting that fluxon motion is responsible for nonlinearity at high reduced temperature.  The anisotropy of YBCO is about 6 [11], whereas for TBCCO it is about 104[12], meaning that the three dimensional transition temperature for TBCCO is nearly coincident with $T_C$ (within 1 K of the critical temperature, $T_C$), whereas it is around 78 K for YBCO.  So, in our measurements from 80 K to $T_C$, YBCO is a 3D superconductor and TBCCO is a 2D superconductor.  In this work, we examine the even and the odd order microwave nonlinearity of YBCO and TBCCO straight-line and folded hairpin resonators at high reduced temperature (t>0.8).

The INFM group in Palermo[13,14] reported peaks in the intrinsic $2^{nd}$ and $3^{rd}$ harmonic response of bulk single crystals just below $T_C$.  To explain harmonic generation in the Meissner state, they used order parameter perturbation in the two-fluid model by considering the response when the normal fluid density is perturbed by the microwave current.  Their model was consistent with an observed slope of 2 in the $3^{rd}$ harmonic power versus input power.  Large peaks occurred in the range of 0.5 K to 1.0 K below $T_C$, and no harmonics from the bulk single crystals were observed above $T_C$.

Lee, *et al.*[15] found peaks at $T_C$ in both the $2^{nd}$ and the $3^{rd}$ harmonics of polycrystalline YBCO thin films (100 nm to 200 nm thick) with the nonlinearity dropping off as the temperature rises above $T_C$.  This cannot be attributed to the Gorter-Casimir model since above $T_C$ the normal fluid density is a constant that is not modulated by the microwave current.  Lee proposed modifying the nonlinear Meissner effect to include a spontaneous time-reversal symmetry breaking (TRSB) current with an onset at $T_C$ and proposed that the best measure of the degree of TRSB is the ratio of $2^{nd}$ order nonlinearity to $3^{rd}$ order nonlinearity $J_2/J_3$ where $J_2$ and $J_3$ are the current densities which occur at the $2^{nd}$ and $3^{rd}$ order nonlinearity frequencies, respectively.

Manuscript received October 9, 2012. This work was supported in part by the U.S. National Science Foundation under Grants DMR-1206149, PHY-0963317 and PHY/DMR-1004811.  Support was also received from the Research Corporation for Science Advancement through a Cottrell College Science Award.  We also acknowledge having received support from Mesaplexx, LTD PTY of Queensland, Australia.
The authors are with the Physics Department, Hope College, Holland, MI 49423, USA (corresponding author phone: 616-395-7507; fax: 616-395-7123; e-mail: remillard@hope.edu).

The above results were obtained on single crystals, presumably free of granularity and Josephson fluxons in the case of INFM, and on unpatterned thin films without contributions from Abrikosov fluxon nucleating edge effects in the case of Lee, *et al.* Zhuravel, *et al.*[16] used a laser scanning microscope to examine patterned YBCO thin film resonators on LAO substrates and found a significantly enhanced 3$^{rd}$ order nonlinearity in microstrips with sloped edges than with vertical edges, as was predicted in the theoretical work of Dahm and Scalapino[17].

Using patterned microwave resonators, this work proceeds along the lines of Zhuravel, *et al.* and Zhao, *et al.*, but incorporates the 2$^{nd}$ order nonlinearity as was done by Lee, *et al.* and INFM. We describe 2$^{nd}$ and 3$^{rd}$ order nonlinearity of microstrip superconducting YBCO or TBCCO thin film resonators with vertical edges on LAO substrates at temperatures below to just above $T_C$. The even and odd order intermodulation distortion (IMD) were both measured at the resonant frequency of the patterned devices using the three-tone intermodulation distortion method developed by our group[18] and based on IMD measurement methods more suitable for use in the wireless industry[19]. Referring to Fig. 1, three probes are introduced into the space above the microstrip device. Probe 1 drives the device with a steady signal at the resonant frequency, $f_r$, for example 850 MHz. Probe 2, ("Probe" in Fig. 1) which is movable, brings in two probing signals at low frequency, e.g. $f_1$=150 KHz and $f_2$=250 KHz. Probe 3 detects the IMD, which is in the passband of the resonant device at $f_r+f_1$ for 2$^{nd}$ order and at, among other frequencies, $f_r+(f_2-f_1)$ for 3$^{rd}$ order. If the tests are performed near $T_C$, the resonator Q is low (<2,000) and both $f_r$ and the IMD tones are essentially at the resonant frequency of the device. The synchronous excitation and IMD generation makes it possible to determine the surface current density, $\hat{n} \times \vec{H}$ in units of A/m, associated with the nonlinearity tone as described in [18]. $\vec{H}$ is the magnetic field at the surface and $\hat{n}$ is the normal to the superconducting film. Also, because of the off-resonance of the probing signals, the IMD is generated locally at the vicinity of Probe 2 [20].

## II. RESULTS

### A. *2$^{nd}$ and 3$^{rd}$ Order IMD of Folded Hairpin Resonators*

The temperature dependence of the 2$^{nd}$ and 3$^{rd}$ order IMD for the same straight-line TBCCO/LAO sample pictured in Fig. 1 was previously reported[21]. A sharp peak in both orders, especially the 3$^{rd}$ order, was observed just below $T_C$. In the current work, two identical folded hairpin shaped resonators shown in the inset of Fig. 2, one made from TBCCO/LAO and the other from YBCO/LAO, were tested near $T_C$ also revealing peaks at $T_C$, though less pronounced. The approximately 400 nm thick films were both on 0.5 mm thick substrates. The lines were 240 μm wide and the entire structure was about 10 mm long and about 3 mm wide. The YBCO sample had $T_C$=87.6K and the TBCCO sample had $T_C$=104.5K, so both were moderately under-doped. Several features are seen in all data. (1) Both 2$^{nd}$ and 3$^{rd}$ order IMD exhibit a slope versus input power in decibels of 1:1. This is expected because three tones are incident and only one is being varied. (2) The IMD surface current density is seen, at least at higher temperatures, to saturate by a carrier surface current density level of about 10 A/m. At lower temperatures, saturation is not reached in the measurement range. (3) For both YBCO and TBCCO, IMD is detected at a temperature up to about 1 K above that at which the peak vanishes.

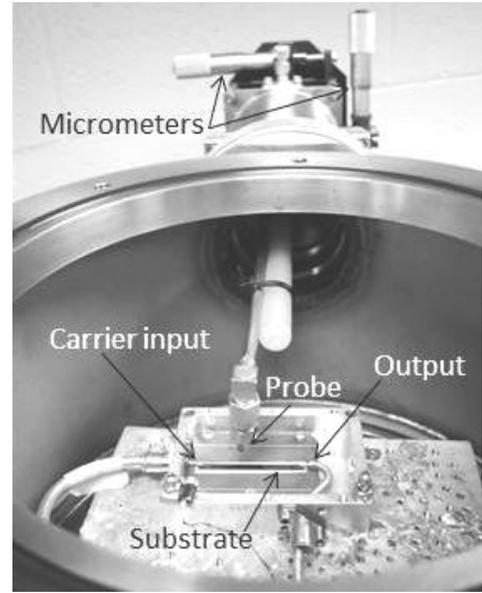

**Fig. 1.** The microwave nonlinearity probe, using LakeShore model MMS-07 Micromanipulated Stage, with a straight HTS transmission line sample. The micrometers are used to position the probe within microns of the patterned line on the substrate.

Fig. 2 shows the temperature dependence of 2$^{nd}$ and 3$^{rd}$ order IMD for a probe located at the electrical center (equidistant from each end) of the folded hairpin resonator. The in-band, or drive, signal was held constant at 4 A/m. The peak at $T_C$ that was so evident in earlier results is also seen in these data. Both samples exhibited a peak in the 3$^{rd}$ order IMD just below $T_C$, although only the TBCCO continues to have high 3$^{rd}$ order as temperature drops. This TBCCO sample also did not exhibit a narrow $T_C$ peak in the 2$^{nd}$ order, although the YBCO sample did, indicating under-doping [15], especially in the TBCCO. Both IMD orders are stronger for the TBCCO at and slightly above $T_C$, perhaps a consequence of superconducting order parameter in the pseudogap state [15], or of the material two-dimensionality enabling quantum mechanical fluctuations above $T_C$ [22]. Experiments with higher probe frequencies will soon be used to rule out a thermal cause of the above $T_C$ nonlinearity.

Keeping in mind that the two resonators were of the same design, and both had the same substrate material (LAO), the often expected result that YBCO produces less 3$^{rd}$ order IMD than TBCCO is validated in Fig. 2. The 2$^{nd}$ order IMD seen in Fig. 2 however is strikingly identical between the YBCO and the TBCCO samples. Unlike the TBCCO sample, the 3$^{rd}$ order IMD drops dramatically as temperature drops for the YBCO sample becoming undetectable below 82K, or a reduced temperature *t* of 0.93, while the 2$^{nd}$ order IMD remains strong for both samples as temperature is reduced, suggesting the possibility that YBCO exhibits a greater degree of TRSB below $T_c$ than does the more anisotropic TBCCO.

The role of fluxons in the two orders of IMD can be investigated using a static magnetic field applied parallel to the film surface. Fig. 3 shows the 2nd and 3rd order IMD of the TBCCO film 2.3K below $T_C$ at 102.2K with the driving surface current density held constant at 2 A/m. The 2nd order IMD is seen to drop by a factor of two as the static field parallel to the HTS surface is raised from zero to 30 Gauss. When the static field is reduced back to zero, the 2nd order IMD traces out a hysteresis loop finally returning approximately to its original, Earth's field value. As also seen from order parameter suppression on single crystals near $T_C$ [14], the presence of Abrikosov fluxons influences the 2nd order IMD by reducing it. The presence of 2nd order nonlinearity in zero field cooled thin film samples points to the contribution of Josephson fluxons, or Abrikosov-Josephson fluxons, in 2nd order nonlinearity, as found using engineered defects [23]. Very little effect is seen when a field is swept over the same range perpendicular to the film. The 3rd order IMD is found to be very weakly effected by the static magnetic field near $T_C$. Hysteresis in the 2nd order, combined with persistence of the 2nd order to lower temperature, leaves open the question about the role of magnetic fluxons on the nonlinearity current at lower temperatures, which will be a subject of future study.

*B. Scanning Capability*

Because three-tone IMD is a local measurement, it is possible to search out centers of high nonlinearity generation within a patterned sample. These centers can be expected at material defects[24] and at high current bends in transmission lines[25]. To prove the concept of three-tone IMD scanning, the measurement reported here was done using a straight-line TBCCO/LAO resonator. The resonant line was 31 mm long and 150 μm wide with a fundamental resonance at 1.28 GHz. 5 mm on each end had gold coatings for optional wire bonding, which were not employed in this sample assembly. The sample was measured at about one Kelvin below $T_C$ in order to maximize the level of observed IMD. The unloaded Q at low power at this high temperature was about 1,100.

Simulations using HFSS show that the peak spatial extent of induced current is about half that of the loop diameter. In these measurements 047 and 086 coaxial cables were used to form the loops, resulting in loop diameters of 600 μm and 1,100 μm, respectively. Newer probes will use cables as small as 023, allowing wire bonded loop diameters as small as 170 μm. Because the micrometers permit a step size, 10 μm, which is smaller than the induced probing current, further improvements in the resolution could be possible with the application of image transformation techniques.

To be meaningful, it is essential that IMD is measured while the local driving field is maintained at the same value. This is accomplished by sweeping through a range of input powers, computing the driving field, and interpolating to the chosen field, which in this case is 2 A/m, because it is below the level where saturation of the IMD begins. The sample, visible in Fig. 1, was modeled using HFSS, from which the best fit to the surface field profile across the sample was determined. The position dependent field $H(x)$ was then written in terms of the dissipated power, which is readily determined from the loaded Q and the coupling coefficients[26].

The 2nd and 3rd order IMD varies as the probe is moved from one end of the sample to the other, as shown in Fig. 4. The IMD varies between $1\times10^{-4}$ A/m and $1\times10^{-3}$ A/m for the 2nd order and $1\times10^{-4}$ A/m and $5\times10^{-4}$ A/m for the 3rd order, except at certain locations where both orders increase significantly. The ratio of 2nd order to 3rd order is also shown in Fig. 4 showing that the ratio, which might reveal the extent of time reversal symmetry breaking, also exhibits hot spots.

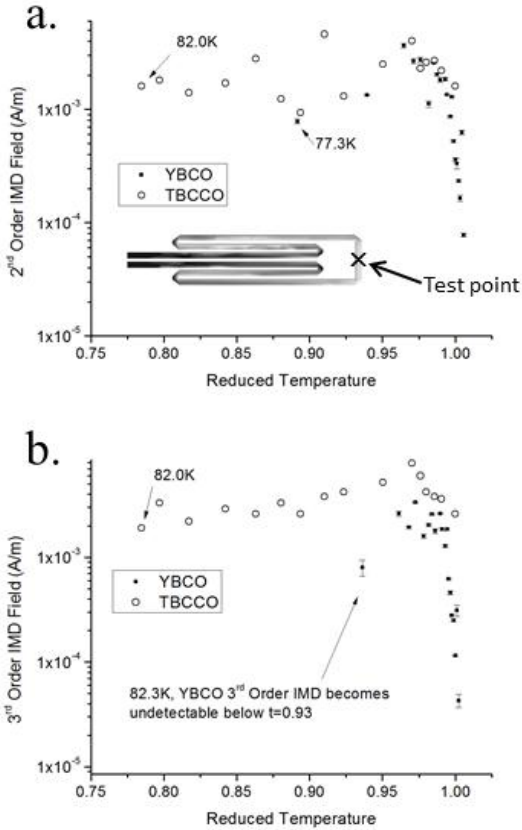

**Fig. 2.** (a) 2nd and (b) 3rd order intermodulation distortion in two identical resonators, one of YBCO and one of TBCCO, at an in-band drive surface current density of 4 A/m. The inset shows the resonance current distribution from HFSS simulation where the lighter tone indicates higher current density.

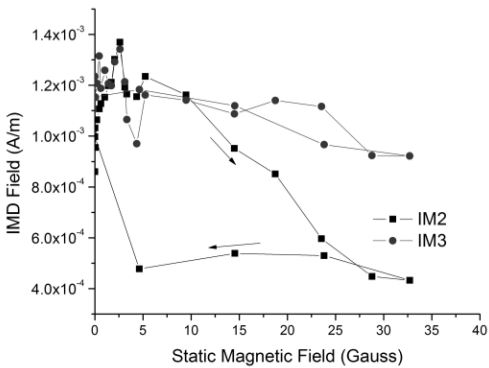

**Fig. 3.** 2nd and the 3rd order IMD depends on DC magnetic field parallel to the film surface, showing hysteresis in the 2nd order IMD, but not in the 3rd order.

The 3rd order IMD exhibits low field saturation at the 2nd order hot spots. Fig. 5 compares Points *f* (x=15 mm) and *de* (x=13 mm). Point *f* has a sharp 2nd order hot spot, and Point *de* is in the neighboring flat region as seen in Fig. 4. A similar situation occurs at the other 2nd order hot spots and their neighboring regions. Away from the hot spot, the 3rd order continues to increase to the limit of the measurement with considerably less reduction in slope at high field. When occurring along with saturation in the device transfer function, harmonic balance effectively models the saturation[27]. The saturation in Fig. 5 is complete, and even exhibits a maximum, although the power transmitted through the device is barely attenuated. This has been shown to result from grain decoupling in harmonic measurements[28], which may yet be established as the cause here too.

Film defects were not evident in optical and SEM exams, nor was a deviation in the film's stoichiometry seen in the EDS analysis, leaving the physical origin of these 2nd order hot spots unknown. There is a prediction, attributed to Varma[15,29], that TRSB domains on the dimensional order of 100 μm are to be expected in underdoped cuprates, consistent with this observation, and the model leading to this expectation has recently been tested by ARPES experiments[30]. Experiments now underway involving temperature dependence will differentiate between TRSB and material inhomogeneity. Irreproducibility on thermal cycling through $T_C$ would provide evidence of local TRSB hot spots. Spatially dependent $T_C$ would provide evidence of material inhomogeneity probably due to spatially varying oxygen, and hence carrier doping, levels. We will also investigate the local nonlinearity generated by engineered microchannel defects which can now be introduced into the microstrip line by such methods as atomic force microscope oxidation[31] and heavy ion beam milling[32].

have demonstrated the capability of this technique to examine the contribution of fluxon creation and annihilation to even order nonlinearity. Continued efforts focus on the probe frequency and the effect of static magnetic fields down to about 50 K, and on probing of more complicated structures.

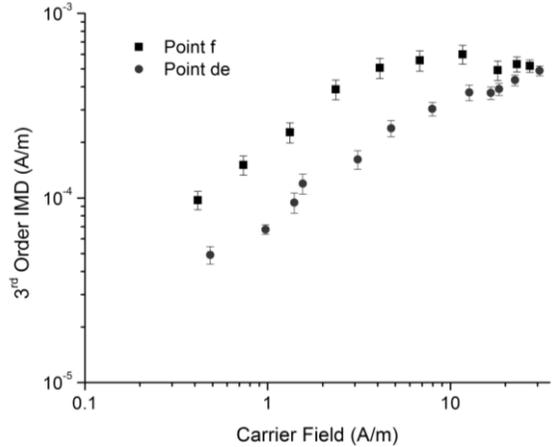

**Fig. 5.** 3rd order IMD over the available range of carrier field at (Point *f* at *x*=15 mm in Fig.4) and away from (Point *de* at *x*=13 mm in Fig. 4) a local 2nd order IMD hot spot.

ACKNOWLEDGMENT

Engineering of the probe station was done by David Daugherty at Hope College. The resonator in Figure 2 was designed and fabricated at Spectral Solutions, Inc. We had useful discussions with Steven Anlage from the University of Maryland and with Sheng-Chiang Lee of Mercer University. The two anonymous reviewers' thought provoking critiques improved this paper and the continued direction of this work.

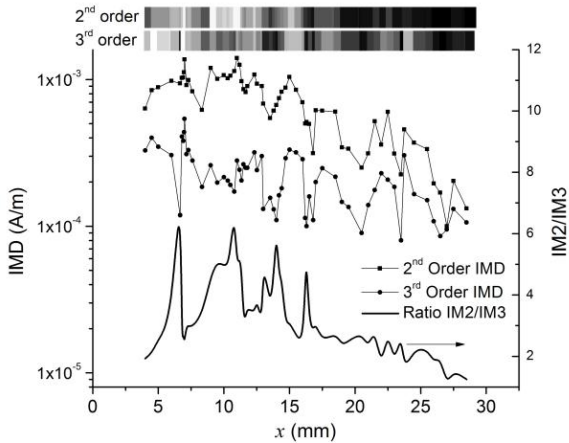

**Fig. 4.** The 2nd (top curve) and 3rd (center curve) order IMD mapped across a straight-line TBCCO resonator at 1 K below $T_C$, along with their ratio (bottom curve). The grey-scale bands on the top indicate high (white) to low (black) IMD.

## III. CONCLUSIONS

Synchronously measured even and odd order nonlinearity can be mapped in superconducting resonant devices revealing the location of 2nd order, and possibly TRSB, hot spots. We